\documentclass[aps,prd,twocolumn,superscriptaddress,showpacs,bibnotes]{revtex4-1}
\usepackage{graphicx}
\usepackage{amsmath}

\begin{document}


\title{Search for a Lorentz-violating sidereal signal with atmospheric neutrinos in IceCube}

\affiliation{III. Physikalisches Institut, RWTH Aachen University, D-52056 Aachen, Germany}
\affiliation{Dept.~of Physics and Astronomy, University of Alabama, Tuscaloosa, AL 35487, USA}
\affiliation{Dept.~of Physics and Astronomy, University of Alaska Anchorage, 3211 Providence Dr., Anchorage, AK 99508, USA}
\affiliation{CTSPS, Clark-Atlanta University, Atlanta, GA 30314, USA}
\affiliation{School of Physics and Center for Relativistic Astrophysics, Georgia Institute of Technology, Atlanta, GA 30332, USA}
\affiliation{Dept.~of Physics, Southern University, Baton Rouge, LA 70813, USA}
\affiliation{Dept.~of Physics, University of California, Berkeley, CA 94720, USA}
\affiliation{Lawrence Berkeley National Laboratory, Berkeley, CA 94720, USA}
\affiliation{Institut f\"ur Physik, Humboldt-Universit\"at zu Berlin, D-12489 Berlin, Germany}
\affiliation{Fakult\"at f\"ur Physik \& Astronomie, Ruhr-Universit\"at Bochum, D-44780 Bochum, Germany}
\affiliation{Physikalisches Institut, Universit\"at Bonn, Nussallee 12, D-53115 Bonn, Germany}
\affiliation{Dept.~of Physics, University of the West Indies, Cave Hill Campus, Bridgetown BB11000, Barbados}
\affiliation{Universit\'e Libre de Bruxelles, Science Faculty CP230, B-1050 Brussels, Belgium}
\affiliation{Vrije Universiteit Brussel, Dienst ELEM, B-1050 Brussels, Belgium}
\affiliation{Dept.~of Physics, Chiba University, Chiba 263-8522, Japan}
\affiliation{Dept.~of Physics and Astronomy, University of Canterbury, Private Bag 4800, Christchurch, New Zealand}
\affiliation{Dept.~of Physics, University of Maryland, College Park, MD 20742, USA}
\affiliation{Dept.~of Physics and Center for Cosmology and Astro-Particle Physics, Ohio State University, Columbus, OH 43210, USA}
\affiliation{Dept.~of Astronomy, Ohio State University, Columbus, OH 43210, USA}
\affiliation{Dept.~of Physics, TU Dortmund University, D-44221 Dortmund, Germany}
\affiliation{Dept.~of Physics, University of Alberta, Edmonton, Alberta, Canada T6G 2G7}
\affiliation{Dept.~of Subatomic and Radiation Physics, University of Gent, B-9000 Gent, Belgium}
\affiliation{Max-Planck-Institut f\"ur Kernphysik, D-69177 Heidelberg, Germany}
\affiliation{Dept.~of Physics and Astronomy, University of California, Irvine, CA 92697, USA}
\affiliation{Laboratory for High Energy Physics, \'Ecole Polytechnique F\'ed\'erale, CH-1015 Lausanne, Switzerland}
\affiliation{Dept.~of Physics and Astronomy, University of Kansas, Lawrence, KS 66045, USA}
\affiliation{Dept.~of Astronomy, University of Wisconsin, Madison, WI 53706, USA}
\affiliation{Dept.~of Physics, University of Wisconsin, Madison, WI 53706, USA}
\affiliation{Institute of Physics, University of Mainz, Staudinger Weg 7, D-55099 Mainz, Germany}
\affiliation{Universit\'e de Mons, 7000 Mons, Belgium}
\affiliation{Bartol Research Institute and Department of Physics and Astronomy, University of Delaware, Newark, DE 19716, USA}
\affiliation{Dept.~of Physics, University of Oxford, 1 Keble Road, Oxford OX1 3NP, UK}
\affiliation{Dept.~of Physics, University of Wisconsin, River Falls, WI 54022, USA}
\affiliation{Oskar Klein Centre and Dept.~of Physics, Stockholm University, SE-10691 Stockholm, Sweden}
\affiliation{Dept.~of Astronomy and Astrophysics, Pennsylvania State University, University Park, PA 16802, USA}
\affiliation{Dept.~of Physics, Pennsylvania State University, University Park, PA 16802, USA}
\affiliation{Dept.~of Physics and Astronomy, Uppsala University, Box 516, S-75120 Uppsala, Sweden}
\affiliation{Dept.~of Physics and Astronomy, Utrecht University/SRON, NL-3584 CC Utrecht, The Netherlands}
\affiliation{Dept.~of Physics, University of Wuppertal, D-42119 Wuppertal, Germany}
\affiliation{DESY, D-15735 Zeuthen, Germany}

\author{R.~Abbasi}
\affiliation{Dept.~of Physics, University of Wisconsin, Madison, WI 53706, USA}
\author{Y.~Abdou}
\affiliation{Dept.~of Subatomic and Radiation Physics, University of Gent, B-9000 Gent, Belgium}
\author{T.~Abu-Zayyad}
\affiliation{Dept.~of Physics, University of Wisconsin, River Falls, WI 54022, USA}
\author{J.~Adams}
\affiliation{Dept.~of Physics and Astronomy, University of Canterbury, Private Bag 4800, Christchurch, New Zealand}
\author{J.~A.~Aguilar}
\affiliation{Dept.~of Physics, University of Wisconsin, Madison, WI 53706, USA}
\author{M.~Ahlers}
\affiliation{Dept.~of Physics, University of Oxford, 1 Keble Road, Oxford OX1 3NP, UK}
\author{K.~Andeen}
\affiliation{Dept.~of Physics, University of Wisconsin, Madison, WI 53706, USA}
\author{J.~Auffenberg}
\affiliation{Dept.~of Physics, University of Wuppertal, D-42119 Wuppertal, Germany}
\author{X.~Bai}
\affiliation{Bartol Research Institute and Department of Physics and Astronomy, University of Delaware, Newark, DE 19716, USA}
\author{M.~Baker}
\affiliation{Dept.~of Physics, University of Wisconsin, Madison, WI 53706, USA}
% ...
\collaboration{IceCube Collaboration}\noaffiliation

\date{\today}

\begin{abstract}
A search for sidereal modulation in the flux of atmospheric muon neutrinos in IceCube was performed.  Such a signal could be an indication of Lorentz-violating physics.   Neutrino oscillation models, derivable from extensions to the Standard Model, allow for neutrino oscillations that depend on the neutrino's direction of propagation.  No such direction-dependent variation was found.   A discrete Fourier transform method was used to constrain the Lorentz and CPT-violating coefficients in one of these models.  Due to the unique high energy reach of IceCube, it was possible to improve constraints on certain Lorentz-violating oscillations by three orders of magnitude with respect to limits set by other experiments.
\end{abstract}


\maketitle


\section{Introduction}
Lorentz invariance and CPT symmetry, which combines charge conjugation (C), coordinate reflection (P), and time reversal (T), are fundamental symmetries of quantum field theory (QFT).  To date, no experimental evidence for a violation of either symmetry has been observed, despite a wide variety of experimental investigations \cite{lrr,tables,cpt}.  However, it remains worthwhile to continue to test these fundamental symmetries, with different experiments and different types of particles, at higher energy scales or with improved sensitivity.  Observation of a violation of one of these symmetries would be an indication of new physics, and possibly point the way towards a unified theory or a theory of quantum gravity.

There is reasonable motivation to expect that Lorentz invariance and CPT symmetry do not hold all the way to the Planck scale ($M_P  \approx 10^{19} $ GeV), due to a discrete structure of spacetime or interactions with a spacetime foam \cite{hawking,samuel}, for example.  Neutrinos are sensitive probes of possible low energy effects of the breaking of these symmetries, because they have very high Lorentz factors and they do not interact by the strong or electromagnetic forces.  Signatures of Lorentz and CPT-violating processes in the neutrino sector may include oscillations with unique energy dependencies, direction-dependent oscillations that violate rotational invariance, or deviations from the anticipated behavior based on the ${L \mathord{\left/
 {\vphantom {L E}} \right.
 \kern-\nulldelimiterspace} E}$ ratio of the experiment \cite{mewes}.

The Standard Model (SM) of particle physics is believed to be the low-energy limit of a more fundamental theory.  Such an extension of the SM is typically assumed to unite QFT and General Relativity at the Planck scale, and to provide a coherent theory of quantum gravity.  To look for signatures of quantum gravity without this fundamental theory, a phenomenological description is necessary.  The Standard Model Extension (SME) \cite{colladay} is an effective-field-theory framework that provides such a phenomenological description at experimentally accessible energies, and has guided numerous searches for possible signatures of Lorentz invariance violation and CPT violation \cite{tables}.

The IceCube detector \cite{karle}, located at the South Pole, is designed for detecting astrophysical neutrinos of all three flavors.  Due to its unique size, it has an unprecedented event rate for high energy atmospheric neutrinos.  Data taken while IceCube operated in a partially-completed, 40-string configuration, was used to search for a periodic variation as a function of right ascension, a possible consequence of a Lorentz-violating preferred frame.  A discrete Fourier transform (DFT) method was used to constrain Lorentz and CPT violating coefficients in the SME, in the context of a direction-dependent neutrino oscillation model that violates rotational invariance.

\section{The Vector Model}
The SME adds to the SM Lagrangian all terms that can be constructed with SM and gravitational fields, but that may also violate Lorentz or CPT symmetries.  The coefficients for these processes have Lorentz indices and represent background tensor fields.  Physically observable phenomena depend on contractions between these tensorial coefficients and the particle momentum.  A subset of the SME, known as the ``minimal'' SME \cite{mewes}, includes all observer-independent, renormalizable, Lorentz and CPT-violating processes.  Energy and momentum are still conserved, and spin-statistics and gauge invariance are preserved.  Right-handed neutrinos are still assumed to decouple and remain undetectable.  Neutrino masses are treated the same as in the SM.

The effective Hamiltonian from the minimal SME, describing Lorentz-violating oscillations between neutrino flavor states $a$ and $b$, is \cite{mewes}
\begin{equation}
\left(h_\mathrm{eff}\right)_{ab}  = \frac{1}{E}\left[ {\left( {a_L } \right)^\mu _{ab} p_\mu   - \left( {c_L } \right)^{\mu \nu } _{ab} p_\mu  p_\nu  } \right],
\label{sme_nu}
\end{equation}
where $E$ is the neutrino energy, $p_ {\mu}$ the neutrino four-momentum, and $\mu$ and $\nu$ are Lorentz indices.  The coefficients $\left( {a_L } \right)^\mu _{ab}$ have mass dimension 1 and lead to Lorentz-violating and CPT-violating interactions.  The coefficients $\left( {c_L } \right)^{\mu \nu} _{ab}$ have mass dimension 0 and lead to Lorentz-violating interactions.  After some approximations applicable to the length and energy scale of atmospheric neutrinos, as well as some assumptions about which components of the interaction tensors are non-zero, a subset model known as the ``vector model'' can be derived \cite{mewes}.  The vector model is convenient for studying possible sidereal variations in the atmospheric neutrino flux.

In the vector model, the only non-zero components of the interaction tensors in Eqn.~\ref{sme_nu} are $\left( {a_L } \right)_{\mu \tau }^X ,\left( {a_L } \right)_{\mu \tau }^Y ,\left( {c_L } \right)_{\mu \tau }^{TX} ,{\rm{ and }}\left( {c_L } \right)_{\mu \tau }^{TY} $, all assumed to be real.  Only vacuum oscillations between neutrino flavor states $\mu$ and $\tau$ are included.  These assumptions are made in a Sun-centered celestial-equatorial coordinate system.  The z-axis is aligned with the Earth's rotational axis and the x-axis points towards the vernal equinox.  Mass-induced oscillations between $\nu_ \mu$ and $\nu_ \tau$ are not included in the vector model.  However, they were included in the simulation of the expected neutrino flux.

\begin{figure}
\includegraphics[width=3.2in]{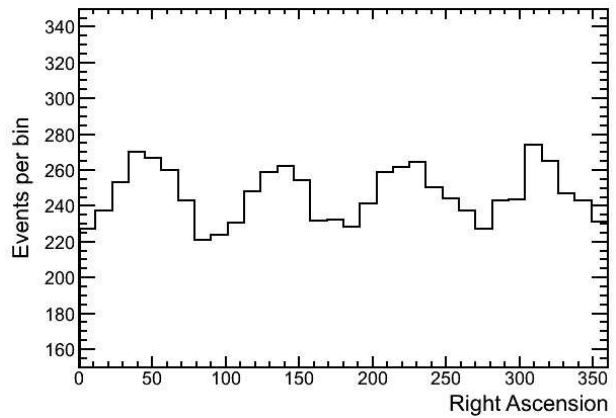}
\caption{Simulation of the sinusoidal signal predicted by Eqn.~\ref{vec_model}, with $a_L ^X = 2 \times 10^{ - 23}$ GeV. }
\label{vm}
\end{figure}

The $\nu _{\mu}$ survival probability is then \cite{mewes}
\begin{equation}
\begin{array}{*{20}c}
   \hfill {P_{\nu _\mu   \to \nu _\mu  }  = } & \hfill {1 - \sin ^2 \left( {L\left[ {\left( A_s \right)_{\mu\tau} \sin \left( {\alpha + \varphi _0 } \right)} \right.} \right.} \\
   \hfill {} & \hfill {\left. {\left. { + \left( A_c \right)_{\mu\tau} \cos \left( {\alpha + \varphi _0 } \right)} \right]} \right)} ,\\
\end{array}
\label{vec_model}
\end{equation}
where $L$ is the propagation distance, $\alpha$ is the neutrino's right ascension, and $\varphi _0$ is an arbitrary phase offset.  Dropping the flavor subscripts, $A_s$ and $A_c$ are defined as
\begin{align}
A_s  &= \hat N^Y \left( {a_L ^X  - 2Ec_L ^{TX} } \right) - \hat N^X \left( {a_L ^Y  - 2Ec_L ^{TY} } \right),\\
A_c  &=  - \hat N^X \left( {a_L ^X  - 2Ec_L ^{TX} } \right) - \hat N^Y \left( {a_L ^Y  - 2Ec_L ^{TY} } \right).
\end{align}
The survival probability for antineutrinos, $P_{\bar \nu _\mu   \to \bar \nu _\mu  }$, is given by changing the sign of the $a _L$ coefficients.  The oscillation probability depends intrinsically on the direction that the neutrino propagates through space, violating rotational invariance.  The $\hat N^{X(Y)}$ are unit propagation vectors for the neutrino:
\begin{align}
 \hat N^X  &= \sin \left( \theta  \right)\cos \left( \varphi  \right), \\ 
 \hat N^Y  &= \sin \left( \theta  \right)\sin \left( \varphi  \right),
\end{align}
where $\theta  = {\pi  \mathord{\left/
 {\vphantom {\pi  2}} \right.
 \kern-\nulldelimiterspace} 2} + \delta $, $\varphi  = \pi  + \alpha $, and $\delta$ is the declination of the incident neutrino.  Figure~\ref{vm} shows an example of the anticipated sinusoidal signal in IceCube, as predicted by Eqn.~\ref{vec_model}, with $a_L ^X = 2 \times 10^{ - 23}$ GeV and the detector configuration and live time discussed in the next section.

\section{The Event Sample}
When completed in 2011, IceCube \cite{karle,str21} will consist of 86 strings.  Each string includes 60 digital optical modules (DOM), for a total of 5,160 DOMs.  A DOM is a single photomultiplier tube and associated electronics in a glass pressure sphere \cite{pmt_paper}.  The instrumented part of the array extends from 1450 m to 2450 m below the surface of the ice.  Horizontally, 78 of the strings are 125 m apart and spread out in a triangular pattern over a square kilometer.  Vertical DOM spacing is a uniform 17 m for these 78 strings.  A subset of the detector, known as ``DeepCore'', consists of 8 specialized and closely spaced strings of sensors located around the center IceCube string.

This analysis used data from 359 days of live time while operating in a 40-string configuration, from April 2008, to May 2009.  No DeepCore strings had been installed at that time.  The event sample is a subset of the data used for an unfolding of the atmospheric muon neutrino spectrum \cite{unfold}.  Triggering, filtering, and background rejection are discussed in detail in \cite{unfold}.  The 40-string detector was roughly twice as long in one horizontal direction as in the other.  However, this azimuthal dependence of the detector shape conveniently canceled out due to the sidereal rotation of the Earth (and thus, the detector around its vertical axis).  

IceCube detects the Cherenkov radiation from charged particles produced in charged current (CC) and neutral current (NC) interactions between incident neutrinos and nucleons in the ice.  If the incident neutrino is a $\nu_ \mu$ or $\bar \nu_ \mu$, a muon or antimuon is produced and undergoes radiative energy losses as it propagates, creating additional Cherenkov radiation.  The muon directions are reconstructed from records of photon arrival times at DOMs participating in each event.  The mean angular deviation between the direction of the parent neutrino and the muon is less than a degree for the energy range of this analysis.  Additionally, simulation and reconstruction studies indicate that muon angular resolution is typically between $0.5^\circ $ and $1^\circ $, depending on the angle of incidence and the muon energy.  Hence, the reconstructed muon direction provides a good estimate for the neutrino direction.

Background events in the data were down-going atmospheric muons, or coincident muons, that were reconstructed as up-going events.  Rejection of this background was done in several stages, beginning with triggering and local coincidence checks on the DOMs \cite{daq}, and software-based filtering at the South Pole \cite{unfold}.  Then, before more computationally intensive reconstructions were performed during off-line processing, unusable events were removed by selection cuts based on zenith angle and track quality parameters.  Finally, using boosted decision trees (BDT) \cite{tmva}, we obtained a sample of 7882 muon neutrino events in the zenith range $97^ \circ$ to $120^ \circ  $, with negligible background.  Background contamination was estimated to be less than $1\%$, based on testing the BDTs with simulated atmospheric muon and neutrino data sets.  This value was then verified by comparing the data passing rate as a function of BDT cut value to the predicted rate from atmospheric muon and neutrino simulation.  These event selection cuts also eliminated localized events from electromagnetic showers induced by $\nu _e$ CC interactions and hadronic showers due to NC interactions.

As discussed in \cite{unfold}, there was a statistically significant excess of events in data (or deficit in simulation) in the zenith region $90^ \circ$ to $97^ \circ  $, the origin of which could not be verified at the time of the analysis.  Hence, that region was not used.  The vector model we adopted assumes that only real components belonging to the plane perpendicular to the Earth's axis are non-zero and ignores any coupling between the z-component of the neutrino momentum and the Lorentz violating coefficients of the SME \cite{mewes}.  By considering only events in the zenith region from $97^ \circ$ to $120^ \circ  $, where the x and y components of the neutrino momenta dominate, the impact of this arbitrary assumption on how the Lorentz violating field is aligned with respect to our preferred coordinate system is minimized.

\begin{figure}
\includegraphics[width=3.2in]{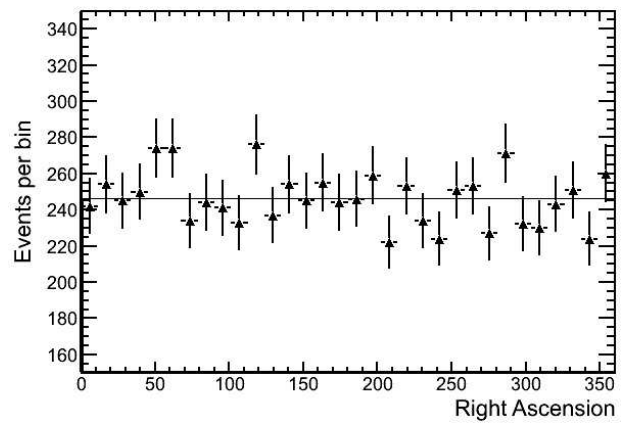}
\caption{RA distribution of events in data.  Vertical error bars are statistical uncertainty only.  Fluctuations in the data, above and below the mean (horizontal line), are consistent with statistical variations.  $\chi ^{2}$ per bin for a straight-line fit to the mean is 0.9.}
\label{ra}
\end{figure}

Figure~\ref{ra} shows the distribution in right ascension (RA) for events in the data.  This histogram has 32 bins from 0 to $360^ \circ$ in RA, the same binning that was used to compute power spectral densities with DFTs as discussed in the next section.  We estimate from simulation that about $90\%$ of the events are from atmospheric neutrinos in the energy range 200 GeV to 13 TeV, and $99\%$ from atmospheric neutrinos in the energy range 100 GeV to 55 TeV.

\section{Methodology and Results}
\label{dft_section}
The DFT analysis methodology was adapted from Ref.~\cite{adamson}, where the MINOS collaboration looked for sidereal variations in the NuMI beam line, using the MINOS near detector.  DFTs were computed using FFTW \cite{fftw} in the ROOT \cite{root} framework.  Under the vector model, the muon neutrino survival probability varies with RA with a modulation frequency of $4\omega _ \oplus  $, where $\omega _ \oplus   = {{2\pi } \mathord{\left/
 {\vphantom {{2\pi } {23{\rm{ }}{\rm{ h }}{\rm{ }}56\min }}} \right.
 \kern-\nulldelimiterspace} {23{\rm{ }}{\rm{ h }}{\rm{ }}56\min }}$ is the Earth's sidereal frequency.  So we are interested in the $n=4$ mode of a DFT of the event rate as a function of RA.  

First, the data were checked for consistency with the hypothesis of no sidereal signal.  For each of 10,000 trials, the RA of all data events were randomized and the power spectral densities (PSD) in modes $n=1$ through $n=4$ of a DFT were computed.  The PSDs for the true data were then computed and compared to these ``noise-only'' distributions.  The data was consistent with no signal in any of the modes.  In particular, the PSDs for data were less than $34\%$ of the noise-only trials for $n=1$, $92\%$ for $n=2$, $31\%$ for $n=3$, and $98\%$ for $n=4$.  

With the absence of a sidereal signal, we were able to set upper limits on the SME coefficients in the vector model.  The flux models of \cite{honda,sarcevic} were assumed for conventional and prompt atmospheric neutrinos.  The predictions for neutrinos from pions and kaons were extended to higher energies by fitting a physics-motivated analytical equation based on energy and zenith angle (\cite{gaisser2} and chapter 7 of \cite{gaisser}), in an overlapping region with the detailed calculations of \cite{honda}.

Normally, we could ignore $\nu _\tau$-induced muons for an atmospheric neutrino analysis.  However, if some of the $\nu _\mu$ and $\bar \nu _\mu$ are oscillating to $\nu _\tau$ and $\bar \nu _\tau$ according to the model that we want to test, then that flux of $\nu _\tau$ and $\bar \nu _\tau$ has to be accounted for.  A $\nu _\tau$ ($\bar \nu _\tau$) could undergo a CC interaction in or near the detector, producing a tau lepton, which could then decay into (among other things) a muon (branching ratio about $17\%$).  Detection of these muons would dampen the signal we are looking for, i.e. a disappearance of muon neutrinos.  This effect was accounted for through toy Monte Carlo studies and $\nu _\tau$ ($\bar \nu _\tau$) simulation.  About $6\%$ of the events lost due to oscillations induced by the $a_ L$ coefficients, and about $9\%$ of the events lost due to oscillations induced by the $c_ L$ coefficients, are recovered.  The difference between the two cases is due to the fact that the mean energy of affected events is higher in the case of the $c_ L$ coefficients, and detector efficiency increases with energy.

In each of 400 toy Monte Carlo (MC) experiments, simulated events were drawn from a distribution matching the energy and zenith dependence of atmospheric neutrinos.  RA's were randomly assigned to the simulated events in each trial.  The physics parameters of the vector model were incrementally increased, and the simulated events re-weighted according to their survival probability under the vector model, until a PSD greater than the 99.87 percentile (equivalent to a 3-sigma threshold) of the PSDs from the 10,000 noise-only toy experiments was obtained.  The values found in each of these 400 trials were then averaged to estimate the sensitivity to a sidereal signal described by the vector model.  These trials were performed independently for each coefficient.  While adjusting one coefficient, the other three were held at zero.

The sensitivity depends on the zenith and energy distribution of atmospheric neutrinos, and is thus affected by uncertainties in these quantities.  Theoretical and experimental uncertainties in the zenith distribution are small, a $3\%$ uncertainty in the predicted ratio of the vertical to horizontal atmospheric neutrino flux \cite{honda}, and angular resolution on the order of a degree \cite{unfold}.  Toy MC experiments with the simulated zenith distribution modified according to these uncertainties showed that the impact on the sensitivity is negligible.

Uncertainties in the energy distribution do not affect sensitivity to oscillations driven by the $a_L^{X(Y)}$ coefficients.  However, uncertainties in the spectral index for atmospheric neutrinos \cite{honda,sarcevic}, and uncertainties in the energy dependence of the detector efficiency, both affect sensitivity to the $c_L^{TX(TY)}$ coefficients.  Uncertainty in the spectral index is primarily due to uncertainty in the energy distribution of the cosmic ray flux \cite{cream,cr_index}.  The uncertainty in the spectral slope of the proton component of the cosmic ray flux is assumed to be $\pm 0.03$, and for the helium component (which makes up roughly $30\%$ of the cosmic rays in the energy region of this analysis) it is assumed to be $\pm 0.07$.  Combining these two factors, after scaling them by the fraction of their contribution to the total flux, leads to an estimated $\pm 0.05$ uncertainty in the spectral index.  Toy MC experiments with simulated atmospheric neutrino events re-weighted to account for this uncertainty in the spectral index showed a $\pm 7\%$ change to the sensitivity for the $c_L^{TX(TY)}$ coefficients.

DOM sensitivity and ice property uncertainties affect the energy dependence of the detector's effective area, and hence the distribution of neutrino energies for events in the data.  Specialized simulated data sets with $\pm 10\%$ enhanced photon populations were used to estimate this uncertainty, in a manner similar to the evaluation of DOM sensitivity and ice property uncertainties for the unfolding analysis discussed in \cite{unfold}.  A $\pm 10\%$ change in the number of photons observed by the DOMs leads to a $\pm 0.05$ change in the spectral index for the energy distribution of detected neutrinos, which in turn leads to a $\pm 7\%$ change in sensitivity for the $c_L^{TX(TY)}$ coefficients.

The uncertainty in DOM sensitivity was estimated to be $\pm 8\%$, based on the measured uncertainty in PMT sensitivity \cite{pmt_paper}.  This was directly scaled to the $\pm 10\%$ change in the number of simulated photons striking the DOMs.  The average photon flux was estimated to change by $\pm 12\%$, due to uncertainties in scattering and absorption, using a diffuse flux approximation \cite{unfold}.  Added in quadrature, DOM sensitivity and ice property uncertainties lead to a $\pm 15\%$ uncertainty in the number of detected photons, and a $\pm 11\%$ uncertainty in sensitivity to the $c_L^{TX(TY)}$ coefficients.

The following upper limits have been set on the SME coefficients, at the 3 sigma level:
\begin{equation}
 a_L^{X}, a_L^{Y} < 1.8 \times 10^{ - 23} {\quad{ }}{\rm{  GeV,}}
\end{equation}
and
\begin{equation}
 c_L^{TX}, c_L^{TY}  < 3.7 \times 10^{ - 27}.
\end{equation}
A net $13\%$ increase ($0.4 \times 10^{ - 27}$) has been added to the upper limit for $c_L^{TX(TY)}$, to account for systematic flux ($7\%$) and detector ($11\%$) uncertainties added in quadrature.

\section{Conclusion}
We have found no sidereal variation in the atmospheric muon neutrino event rate in IceCube.  In the context of the SME, we found no evidence for a violation of Lorentz or CPT symmetries due to a preferred reference frame.  The LSND \cite{auerbach} and MINOS \cite{adamson,adamson2} Collaborations also did not see sidereal variations in the number of neutrinos detected from their respective beam lines.  With their far detector \cite{adamson2}, MINOS found $a_L^X < 5.9 \times 10^{ - 23}$ GeV and $a_L^Y < 6.1 \times 10^{ - 23} $ GeV, $c_L^{TX} $ and $c_L^{TY} < 0.5 \times 10^{ - 23} $.  Our results of $a_L^{X}, a_L^{Y} < 1.8 \times 10^{ - 23} {\quad{ }}{\rm{  GeV,}}$ and $c_L^{TX}, c_L^{TY}  < 3.7 \times 10^{ - 27}$, have improved upon these limits by a factor of three for the $a_L^{X(Y)}$ coefficients and by three orders of magnitude for the $c_L^{TX(TY)}$ coefficients, due to the long baseline of atmospheric neutrinos and the high energy reach of IceCube.

\begin{acknowledgments}

We acknowledge the support from the following agencies:
U.S. National Science Foundation-Office of Polar Programs,
U.S. National Science Foundation-Physics Division,
University of Wisconsin Alumni Research Foundation,
the Grid Laboratory Of Wisconsin (GLOW) grid infrastructure at the University of Wisconsin - Madison, the Open Science Grid (OSG) grid infrastructure;
U.S. Department of Energy, and National Energy Research Scientific Computing Center,
the Louisiana Optical Network Initiative (LONI) grid computing resources;
National Science and Engineering Research Council of Canada;
Swedish Research Council,
Swedish Polar Research Secretariat,
Swedish National Infrastructure for Computing (SNIC),
and Knut and Alice Wallenberg Foundation, Sweden;
German Ministry for Education and Research (BMBF),
Deutsche Forschungsgemeinschaft (DFG),
Research Department of Plasmas with Complex Interactions (Bochum), Germany;
Fund for Scientific Research (FNRS-FWO),
FWO Odysseus programme,
Flanders Institute to encourage scientific and technological research in industry (IWT),
Belgian Federal Science Policy Office (Belspo);
University of Oxford, United Kingdom;
Marsden Fund, New Zealand;
Japan Society for Promotion of Science (JSPS);
the Swiss National Science Foundation (SNSF), Switzerland;
A.~Gro{\ss} acknowledges support by the EU Marie Curie OIF Program;
J.~P.~Rodrigues acknowledges support by the Capes Foundation, Ministry of Education of Brazil.

\end{acknowledgments}


\end{document}